\documentclass[reprint,twocolumn,superscriptaddress,showkeys,
nofootinbib,notitlepage,amsmath,amssymb,floatfix]{revtex4-2}

\usepackage{latexsym,amsmath,amssymb,lmodern,float,url}
\usepackage{natbib}
\usepackage{color}
\usepackage{multirow}
\usepackage{bm}
\usepackage{array}
\usepackage[normalem]{ulem}
\usepackage{cleveref}

\usepackage{mathtools}

\def\de{\delta^{\vphantom{1}}}
\def\bde{{\bar\delta}}

\def\h3{{\displaystyle{\frac 3 2}}}

\newcommand{\bbar}{\overline}

\def\schro{Schr\"odinger~}

\begin{document}
\title{$T_{cc}$ in the Diabatic Diquark Model: Effects of $D^* D$ Isospin}
\author{Richard F. Lebed}
\email{Richard.Lebed@asu.edu}
\author{Steven R. Martinez}
\email{srmart16@asu.edu}
\affiliation{Department of Physics, Arizona State University, Tempe,
AZ 85287, USA}
\date{June, 2024}

\begin{abstract}
$T_{cc}^+$ is an isoscalar 4-quark state with mass lying barely below the $D^{*+} D^0$ threshold, and several times further below the $D^{*0} D^+$ threshold.  It allows both di-meson molecular and elementary diquark-antidiquark $(cc)(\bar u \bar d)$ substructures.  The diabatic generalization of the adiabatic approximation within the Born-Oppenheimer formalism rigorously incorporates the mixing of such elementary eigenstates with states corresponding to two-particle thresholds.  We examine the separate influence of the two $D^* \! D$ isospin channels and find that the influence of $D^{*+} D^0$ is larger than that of $D^{*0} D^+$ but not overwhelmingly so, and that $T_{cc}^+$ contains an $O(10\%)$ $(cc)(\bar u \bar d)$ component.  We then explore the variation of these results if the isospin breaking between the di-meson thresholds is varied, and also the sensitivity of our results to variation of the mixing-potential parameters.
\end{abstract}

\maketitle

\section{Introduction}\label{sec:Intro}

Even among the $>60$ heavy-quark exotics already observed, the state $T_{cc}^+$ discovered by LHCb~\cite{LHCb:2021vvq,LHCb:2021auc} stands out as unique.  First, it is the only exotic state observed to date with open heavy flavor ($cc\bar u \bar d$).  It lies only a few hundred keV below the $D^{*+} D^0$ threshold, and it has the smallest width of any hadron that decays strongly through non-Okubo-Zweig-Iizuka (OZI)-suppressed modes (the only available open-charm channels being $D^+ \! D^0 \pi^0$ and its discovery mode $D^0 D^0 \pi^+$).  Its measured parameters are~\cite{ParticleDataGroup:2022pth}:
\begin{eqnarray}
m_{T_{cc}^+} & = & \ 3874.83 \pm 0.11 \ {\rm MeV} \, , \nonumber \\
m_{T_{cc}^+} - m_{D^{*+}} - m_{D^0} & = & \ \ \ \ -270 \pm \ \hphantom{1} 60 \ {\rm keV} \, , \nonumber \\
\Gamma_{T_{cc}^+} & = & \ \ \ \ \hphantom{+} 410 \pm \ 170 \ {\rm keV} \, .
\label{eq:Tccdata}
\end{eqnarray}
In comparison, the famous $\chi_{c1} (3872)$ [or $X(3872)$] lies at almost the same mass, extremely close to the $\bar D^{*0} \! D^0$ ($\! = \!\! D^{*0} \bar D^0$) threshold, but has a larger width, chiefly due to its hidden-charm content permitting charmonium decays [$J/\psi$, $\chi_{c1}(1P)$, and $\psi(2S)$ modes being observed to date], as well as open-charm decays [$D^0 \bar D^0 \pi^0$ and $\bar D^{*0} D^0$]:
\begin{eqnarray}
m_{\chi_{c1}(3872)} & = & 3871.65 \pm 0.06 \ {\rm MeV} \, , \nonumber \\
m_{\chi_{c1}(3872)} - m_{\bar D^{*0}} - m_{D^0} & = & \ \ \ \ -40 \pm \ \ \ 90 \ {\rm keV} \, , \nonumber \\
\Gamma_{\chi_{c1}(3872)} & = & \ \ \ 1190 \pm \ \ 210 \ {\rm keV} \, .
\label{eq:X3872data}
\end{eqnarray}

In both cases, the isospin partner to the di-meson threshold nearest to the resonance lies somewhat higher:\footnote{In calculating these values, we have used the three best-de\-termined independent mass splittings~\cite{ParticleDataGroup:2022pth} among the $D^{(*)}$ me\-sons: $\left( m_{D^\pm} - m_{D^0} \right) = 4.822(15)~{\rm MeV}$, $\left( m_{D^{*+}} - m_{D^0} \right) = 145.4258(17)~{\rm MeV}$, and $\left( m_{D^{*0}} - m_{D^0} \right) = 142.014(30)~{\rm MeV}$.} 
\begin{eqnarray}
\left( m_{D^{*0}} + m_{D^+} \right) - \left( m_{D^{*+}} + m_{D^0} \right) & = & 1.411 \pm 0.034 \ {\rm MeV} \, , \nonumber \\
\left( m_{D^{*-}} + m_{D^+} \right) - \left( m_{\bar D^{*0}} + m_{D^0} \right) & = & 8.234 \pm 0.034 \ {\rm MeV} \, . \nonumber \\
\label{eq:TccIsospin}
\end{eqnarray}
The difference is significantly smaller for the $T_{cc}$ case than the $\chi_{c1}(3872)$ case, suggesting that the structure of $T_{cc}$ is more strongly influenced than $\chi_{c1}(3872)$ by isospin-dependent di-meson threshold effects, while $\chi_{c1}(3872)$ has significant coupled-channel charmonium decay effects (indicated by its larger decay width) that are suppressed for $T_{cc}$, a point first clearly stated and explored in Ref.~\cite{Kinugawa:2023fbf}.

Both $T^+_{cc}$ and $\chi_{c1}(3872)$ are isoscalar states; searches for isospin (charge) partners (Refs.~\cite{LHCb:2021auc} and \cite{Aubert:2004zr,Choi:2011fc}, respectively) yield no significant signals.  However, $\chi_{c1}(3872)$ is observed to decay to (indeed, was discovered in~\cite{Choi:2003ue}) the channel $J/\psi \, \pi^+ \pi^-$.  The $I=0$ $\chi_{c1}(3872)$ has $J^{PC} = 1^{++}$~\cite{LHCb:2015jfc} and hence $G \equiv C (-1)^I = +$, while the $I=0$ $J/\psi$ has $J^{PC} = 1^{--}$ and hence $G=-$.  Then, by $C$-conservation in strong decays, the $\pi^+ \pi^-$ pair ($G = +$) must have $C = -$, and hence $I = 1$; thus, $\chi_{c1}(3872)$ exhibits some isospin-violating decay modes.  For $T_{cc}^+$, $J^P = 1^+$ is heavily favored due both to the $S$-wave quantum number of the $D^{*+} D^0$ threshold pair and to the absence of a signal in the $D^+ \! D^0$ channel (since two $0^-$ mesons cannot form a $1^+$ in any partial wave).

While the di-meson pairs $D^{*0} \bar D^0$ and $D^{*+} D^0$ in their $S$ waves naturally provide the $J^P = 1^+$, $I=0$ quantum numbers proven or heavily favored for $\chi_{c1}(3872)$ and $T_{cc}^+$, respectively, these composite quasi-molecular combinations do not represent the only natural substructure capable of explaining the states.  In addition, the diquark combinations $(cq)_{\bar {\bf 3}} (\bar c \bar q)^{\vphantom{X}}_{\bf 3}$ and $(cc)_{\bar {\bf 3}} (\bar u \bar d)^{\vphantom{X}}_{\bf 3}$ (color triplets being the most attractive diquark combination) each naturally produce a spectrum of states that includes $J^P = 1^+$.   These ``elementary'' diquark combinations (in contrast to the composite di-meson components) provide a natural alternative component for the full states.\footnote{And of course, the conventional charmonium state $\chi_{c1}(2P)$ can contribute to $\chi_{c1}(3872)$---an added complication for this state.}

The complete spectrum of purely elementary diquark combinations $(Qq)_{\bar {\bf 3}} (\bar Q \bar q)_{\bf 3}$, where $Q$ denotes a heavy quark, has been studied in the context of the dynamical diquark model~\cite{Brodsky:2015wza,Lebed:2017min,Giron:2019bcs,Giron:2019cfc,Giron:2020fvd,Giron:2020qpb}.  A key ingredient of this model is the presence of configurations in which the diquark states do not always instantaneously reorganize into di-meson $(Q\bar q)(\bar Q q)$ combinations, which in this model is realized through components of the 4-quark configuration for which the relative momentum of the $(Qq)$ and ($\bar Q \bar q$) diquarks is larger than that within either diquark.  Each heavy quark $Q$,$\bar Q$ then serves to nucleate a spatially localized diquark quasiparticle (due to its small Fermi momentum), thus producing a configuration that can be distinguished from a di-meson pair.

In contrast, for $(Q_1 Q_2)(\bar q_1 \bar q_2)$ combinations like those relevant to $T_{cc}$, the mechanism of nucleating the $(\bar q_1 \bar q_2)$ diquark using a heavy quark is no longer available.  In this case, long-established phenomenology~\cite{Jaffe:2004ph}, as well as more recent lattice simulations~\cite{Francis:2021vrr}, assert that the $(\bar u \bar d)_{\bf 3}$ diquark in its ``good'' $S=0$, $L=0$, $I=0$ channel is the most tightly bound of all the possibilities, while the $(cc)_{\bar {\bf 3}}$ diquark (antisymmetric in color) is not only compact due to the small Fermi momentum of the two heavy quarks, but in its ground state has $S=1$ and $L=0$ ({\it i.e.}, is symmetric in spin and space) in order to satisfy Fermi statistics.  Assuming lastly that the ground state of the $(cc)(\bar u \bar d)$ combination also occurs in a relative $S$-wave between the diquarks, then the state has the unique quantum numbers $J^P = 1^+$, $I=0$, exactly as appears to be true for $T_{cc}^+$.

Numerous calculations performing coupled-channel analyses of exotic hadrons appear in the literature, including for $T_{cc}^+$.  Analyses including both molecular and elementary components for $T_{cc}^+$ include Refs.~\cite{Kinugawa:2023fbf,Chen:2021tnn,Dai:2023cyo,Dai:2023kwv,Albaladejo:2023wmv,Chen:2023fgl,Liu:2023ckj,Ortiz-Pacheco:2023ble}.  In this work, we first consider $T_{cc}^+$ as an elementary $(cc)_{\bar {\bf 3}}(\bar u \bar d)_{\bf 3}$ state initially interacting through a color-triplet static potential---{\it i.e.}, using the Born-Oppenheimer (BO) approximation---and then employ the rigorous {\it diabatic\/} formalism~\cite{Baer:2006} that generalizes the adiabatic formalism inherent in the BO approximation, in order to include the effects of coupled di-meson channels on the state.  The diabatic approach was applied for the first time in hadronic physics to treat exotic heavy-quark hadrons as mixtures of quarkonium with di-meson states in Ref.~\cite{Bruschini:2020voj}, and later as mixtures of diquark-antidiquark states with di-meson states in Ref.~\cite{Lebed:2022vks}.  Very recently, the diabatic formalism was generalized~\cite{Bruschini:2023zkb} to incorporate the combined effects of channels with distinct di-meson quantum numbers (in particular, spin).

In the case of $T_{cc}^+$, the most important thresholds are of course the two isospin channels $D^{*+} D^0$ and $D^{*0} D^+$, but a complete analysis would also include their heavy-quark spin partner $D^{*+} D^{*0}$, approximately 140~MeV higher.  In this initial study, only the lower channels are included, as our goal is to determine the separate effect of each of these channels on the structure of the state.  That is, we analyze the effect of not only the nearest $D^{*+} D^0$ threshold, but also the effect of its $D^{*0} D^+$ isospin partner about 1.4~MeV higher.  As noted above, $T_{cc}^+$ provides a cleaner laboratory than $\chi_{c1}(3872)$ for examining the effects of isospin breaking, and we study its consequences both for the observed value of isospin breaking, and also parametrically as this number is varied.

This paper is organized as follows.  In Sec.~\ref{sec:Diquark} we discuss the diquark configuration relevant to $T_{cc}^+$ and other $(Q_1 Q_2) (\bar q_1 \bar q_2)$ states in the BO approximation.  Section~\ref{sec:Diabatic} presents a brief review of the diabatic formalism in its original form for hadrons, while Sec.~\ref{sec:MultiLevel} presents improvements to the formalism designed to incorporate distinct spin and/or flavor di-hadron thresholds.  In Sec.~\ref{sec:Analysis} we analyze the $T_{cc}^+$ system in the diabatic formalism including both $D^{*+} D^0$ and $D^{*0} D^+$ channels, and in Sec.~\ref{sec:Concl} we present our conclusions and indicate directions for further improvements.

\section{Diquark Model for $T_{cc}^+$} \label{sec:Diquark}

As noted in the Introduction, the state $T_{cc}^+$---assuming the confirmation of its $I=0$, $J^P =1^+$ quantum num\-bers---admits two natural substructures to accommodate its valence-quark $c c \bar u \bar d$ content: As a $D^*D$ molecule (possibly with an admixture of $D^* D^*$), and as a bound state of a compact color-$\bar {\bf 3}$ $(cc)$ diquark and a ``good'' (color {\bf 3}, $I=0$, $S=0$) $(\bar u \bar d)$ antidiquark.  While the extreme proximity of the $D^{*+} D^0$ threshold [Eq.~(\ref{eq:Tccdata})] is quite suggestive of this di-meson substructure, it does not rule out the possibility of a significant diquark component.  Moreover, even a small diquark component in the $T_{cc}^+$ state does not indicate that the diquark structure is unimportant; as shown in the diabatic formalism in Refs.~\cite{Lebed:2022vks,Lebed:2023kbm} to be discussed in Sec.~\ref{sec:Diabatic}, a $(cq)(\bar c \bar q)$ diquark structure (where $q$ is a light quark) with $I=0$ can easily serve as a ``seed'' for generating the famous $\chi_{c1} (3872)$ that lies so close to the $\bar D^{*0} \! D^0$ threshold [Eq.~(\ref{eq:X3872data})].

Indeed, the best corroborating evidence for the presence of a significant $(cc) (\bar u \bar d)$ component with a ``good'' diquark would be the absence of a prominent state in a channel that relies upon a light diquark with ``bad'' quantum numbers, such as $(\bar u \bar d)$ with $S=1$ or $I=1$: {\it i.e.}, $T_{cc}^+$ with $J^P = 2^+$ or a $T_{cc}^{++}$ partner; or of a strange analogue such as the charge-$2$ $(cc)(\bar s \bar s)$.  Since both phenomenology~\cite{Jaffe:2004ph} and recent lattice calculations~\cite{Francis:2021vrr} indicate a ``bad''-to-``good'' light-diquark mass difference of at least 200~MeV, then no additional $T_{cc}$ states much below 4100~MeV should arise under the assumptions of this model.\footnote{For a differing view, see Ref.~\cite{Mutuk:2024vzv}.}  Additional experimental evidence for this conclusion is provided by the Belle Collaboration, which as yet has seen no evidence for $D_s^+ D_s^+$ or $D_s^{*+} D_s^{*+}$ resonances in $\Upsilon(1S)$ or $\Upsilon(2S)$ decays~\cite{Belle:2021kub}.

Another way to visualize the exceptional nature of the state $T_{cc}^+$ is by examining the full ground-state multiplet of tetraquark states in a diquark-antidiquark ($\de$-$\bde$) picture.  Self-conjugate [$(Qq)(\bar Q \bar q)$] states produce the spectrum~\cite{Maiani:2014aja,Lebed:2017min}:
\begin{eqnarray}
J^{PC} = 0^{++}: & \ & X_0 \equiv \left| 0_\de , 0_\bde \right>_0 \,
, \ \ X_0^\prime \equiv \left| 1_\de , 1_\bde \right>_0 \, ,
\nonumber \\
J^{PC} = 1^{++}: & \ & X_1 \equiv \frac{1}{\sqrt 2} \left( \left|
1_\de , 0_\bde \right>_1 \! + \left| 0_\de , 1_\bde \right>_1 \right)
\, ,
\nonumber \\
J^{PC} = 1^{+-}: & \ & \, Z \  \equiv \frac{1}{\sqrt 2} \left( \left|
1_\de , 0_\bde \right>_1 \! - \left| 0_\de , 1_\bde \right>_1 \right)
\, ,
\nonumber \\
& \ & \, Z^\prime \equiv \left| 1_\de , 1_\bde \right>_1 \, ,
\nonumber \\
J^{PC} = 2^{++}: & \ & X_2 \equiv \left| 1_\de , 1_\bde \right>_2 \,
.
\label{eq:Swavediquark}
\end{eqnarray}
Here, the $\de(\bde)$ spin is denoted by $s^{\vphantom\dagger}_{\de}(s_{\bde})$, and the total state spin is designated by the outer subscript.  In the present case, $\de \bde \! = \! (cc)(\bar q_1 \bar q_2)$ states are not self-conju\-gate, and therefore the $C$-parity quantum number is lost.  Furthermore, we have seen that $\de \! = \! (cc)$ in its ground state gives $s_\de \! = \! 1$; and since in addition the ``good'' diquark $\bde = (\bar u \bar d)$ has $s_\bde \! = \! 0$, then all states in Eqs.~(\ref{eq:Swavediquark}) disappear except for the first components of $X_1$ and $Z$, which coalesce to leave a single $J^P \! = 1^+$ state.

 The Introduction also notes that the dynamical diquark model in its original form is not applicable for doubly heavy open-flavor hadrons like $T_{cc}$, because one of the two quasiparticle components (here, $\bde$) lacks a heavy quark, and hence the system has no typical configuration in which the $\de$ and $\bde$ components can be described as spatially separated.  In fact, the most natural diquark picture for $T_{cc}^+$ much more closely resembles hadrocharmonium~\cite{Voloshin:2007dx,Dubynskiy:2008mq}, in which the heavy quarks form a compact nucleus embedded within the light-quark cloud, except that for $T_{cc}^+$ the ``nucleus'' is color-$\bar {\bf 3}$, and $\bde$ is color-{\bf 3}.  Directly using a {\bf 3}-$\bar{\bf 3}$ interaction such as the Cornell potential~\cite{Eichten:1978tg,Eichten:1979ms},
\begin{equation} \label{eq:Cornell}
V(r) = -\frac{a}{r} + b r \, ,
\end{equation}
or a lattice simulation of the potential between two heavy colored sources~\cite{Juge:1997nc,Juge:1999ie,Juge:2002br,Capitani:2018rox} is highly questionable in the current circumstances.  Instead, here we model the state by supposing that the potential of Eq.~(\ref{eq:Cornell}) is valid for pointlike sources, and then treat the diquark $(cc)$ as pointlike, but regard the light diquark $(\bar u \bar d)$ as a sphere with uniform density for $0 < r  < R$, with $R$ given by its root-mean-square radius $\left< r^2 \right>_{(ud)}^{1/2}$ obtained from phenomenology or lattice simulations.  (Of course, one could choose to model $\bde$ using any other specific wave-function profile.) With this modification, $r$ indicates the position of $(cc)$ with respect to the center of the $(\bar u \bar d)$  wave function, and the full potential in this case is then straightforward to compute:
\begin{equation}\label{eq:VforTcc}
V(r) =
\begin{cases}
- \left( a - \frac{b R^2}{5} \right) \frac 1 r + b r , & r \geq R , \\
+\frac{a}{2R^3} \left( r^2 - 3R^2 \right) \\ \hspace{0.5em} - \frac{b}{20R^3} \left( r^4 - 10 R^2 r^2 - 15 R^4 \right) , & r \leq R .
\end{cases}
\end{equation}
This expression, when restricted to $b \! = \! 0$, is the textbook result obtained from applying Gauss' law to a uniform-density sphere in the $1/r$ potential of electrostatics or gravity.  The analysis of the confining ($b \! \neq \! 0$) part of the potential is slightly more complicated, since in that case the potential at a given value of $r$ depends upon the effects of sources both inside and outside the sphere of radius $r$.  One finds, interestingly, that the strength of the Coulomb term decreases outside the sphere for any value of $R$.

The general form for the elementary $\de$-$\bde$ potential used in this work is the same as in Refs.~\cite{Lebed:2022vks,Lebed:2023kbm},
\begin{equation}
\label{eq:sgmapot}
V_{\de \bde}(r)=- \frac{\alpha}{r} + \sigma r + V_0 + m_{\de} +
m_{\bde} \, ,
\end{equation}
and uses lattice-determined values~\cite{Morningstar:2019} of $\alpha,\sigma,$ and $V_0$:
\begin{eqnarray} \label{eq:sgmapotparams}
\alpha & = & 0.053 \ \rm{GeV} \! \cdot \rm{fm}, \nonumber \\
\sigma & = & 1.097 \ \rm{GeV \! /fm}, \nonumber \\
V_0 & = & -0.380 \ \rm{GeV}.
\end{eqnarray}
Of course, the parameters $a,b$ of the original Cornell potential Eq.~(\ref{eq:Cornell}) are directly replaced in the current potential Eq.~(\ref{eq:sgmapot}) with $\alpha,\sigma$, respectively, and the specific value listed for $V_0$ is only taken as a starting point for fits.

%

As for values of diquark masses and $\left< r^2 \right>_{(ud)}^{1/2}$, one may resort to results from phenomenology, QCD sum rules, or lattice calculations.  Values of $m_{(ud)}$ are broadly consistent; listing results from three papers that provide uncertainties, we find $m_{(ud)} = 640 \pm 60$~MeV~\cite{Wang:2011ab} (QCD sum rules); $694 \pm 22$~MeV~\cite{Hess:1998sd} and $690 \pm 47$~MeV~\cite{Bi:2015ifa} (lattice).  For definiteness, we use the last (most recent) of these determinations, which employs an unquenched simulation.  Values of $\left< r^2 \right>_{(ud)}^{1/2}$ are less commonly presented; here we start with the estimate of 0.6~fm presented in a rather recent lattice simulation~\cite{Francis:2021vrr}.  Lastly, values of $m_{(cc)}$ vary widely; here one finds results such as $3510 \pm 350$~MeV~\cite{Esau:2019hqw} (QCD sum rules); $3306.2$~MeV~\cite{Karliner:2017qjm} and $3136 \pm 10$~MeV~\cite{Giron:2020wpx} (phenomenology).  Due to this large spread, here we simply use the measured mass value of $T_{cc}^+$ to fit for $m_{(cc)}$, and then study the results as the parameters $m_{(ud)}$, $\left< r^2 \right>_{(ud)}^{1/2}$, and $m_{(cc)}$ are varied.

Lastly, note that the $(\bar u \bar d)$ diquark itself contains no heavy quarks, but its mass is nevertheless substantially larger than $\Lambda_{\rm QCD}$ (and indeed is about the same as that of a pair of constituent light quarks in a typical phenomenological quark model).  For the purposes of these calculations, we treat $(\bar u \bar d)$ as a heavy---but not pointlike---source in the BO approximation.  In order for the BO formalism to be relevant here, the light degrees of freedom (d.o.f.) associated with the potential $V(r)$ must be able to adjust quickly to changes in the configuration of the heavy sources $(cc)$ and $(\bar u \bar d)$.  We take as evidence for this criterion to be satisfied that typical values of the potential $V(r)$ for the state must be small compared to $m_{(ud)}$, and we find explicitly that this requirement is satisfied in our calculations.

We note that our calculation is certainly not the first to model $T_{cc}^+$ using diquarks, nor even the first to use the BO approximation for such a tetraquark.  For example, Ref.~\cite{Maiani:2022qze} treats $T_{cc}^+$ analogously to a $H_2$ molecule in the BO approximation with all 4 quarks initially dynamical, and performs a variational calculation to obtain the BO potential and hence the eigenvalue spectrum.  In our calculation, the diquarks $(cc)$, $(\bar u \bar d)$ are introduced as quasiparticles, with $(\bar u \bar d)$  having a definite spatial extent comparable to the value noted in Ref.~\cite{Maiani:2022qze}, but the size of $(cc)$ is irrelevant for us.  Furthermore, we do not include hyperfine interactions between the diquarks, since by our argument above the diquarks in $T_{cc}^+$ occur in uniquely specified spin states.  The innovations of this work include the explicit incorporation of di-meson thresholds within the BO approach, as well as the derivation of an interaction [Eq.~(\ref{eq:VforTcc})] that describes interactions between the diquarks when they spatially overlap.

\section{The Diabatic Formalism} \label{sec:Diabatic}

The modeling of an exotic 4-quark hadron solely by a $\de \bde$ pair interacting through a potential $V(r)$, such as in the context of the BO approximation, intrinsically neglects the effects of coupling to di-meson thresholds.  A coupled-channel formalism is clearly required if one wishes to incorporate this important dynamical source.  While coupled-channel calculations are nothing new in the literature, they can often seem rather {\it ad hoc\/} in their implementation.  The BO approximation, however, possesses a rigorous generalization called the {\it diabatic formalism\/} that has become a standard, textbook approach in the context of atomic and molecular physics~\cite{Baer:2006}.  It was first applied in the context of hadronic physics relatively recently~\cite{Bruschini:2020voj}, in order to study the coupling of heavy quarkonium to exotic hadrons with the same $J^{PC}$ quantum numbers.  The first introduction of $\de \bde$ degrees of freedom coupled to the di-meson thresholds followed in Ref.~\cite{Lebed:2022vks}.  The diabatic formalism was generalized to perform direct studies of the scattering amplitudes in which the quarkoniumlike states appear in Ref.~\cite{Bruschini:2021ckr} for $Q\bar Q$  states, and in Ref.~\cite{Lebed:2023kbm} for $\de \bde$ states.  Calculations of mass shifts and strong decay widths induced by the couplings to the di-meson thresholds were investigated for $Q\bar Q$ states in Ref.~\cite{Bruschini:2021cty} and for $\de \bde$ states in Ref.~\cite{Lebed:2024rsi}.  Since the diabatic approach is described in all of these papers, here we present only a brief summary.

One begins with a Hamiltonian for a system of two heavy color sources interacting through light fields:
\begin{equation} 
\label{eq:SepHam}
H=K_{\rm heavy} + H_{\rm light} =
\frac{\mathbf{p}^2}{2 \mu_{\rm heavy}} + H_{\rm light}.
\end{equation}
Here, $H_{\rm light}$ contains both the light-field static energy and the heavy-light interaction.  Defining $\mathbf{r}$ as the separation vector for the heavy-source pair (with corresponding eigenstates $| \mathbf{r} \rangle$) and $|\xi_i(\mathbf{r}) \rangle$ as the $i^{\rm th}$ eigenstate of $H_{\rm light}$ with eigenvalue $E_i$, one may expand the solutions to the corresponding \schro equation as:
\begin{equation} 
\label{eq:AdExp}
|\psi \rangle = \sum_{i} \int d\mathbf{r} \, \tilde \psi_i
(\mathbf{r}) \, |\mathbf{r} \rangle \:
|\xi_i(\mathbf{r}) \rangle .
\end{equation}
While this expansion already suggests the decoupling separation of the BO approximation between heavy and light d.o.f., it still can be used in the general case.  The set $\{ |\xi_i(\mathbf{r}) \rangle \}$ forms a complete, orthonormal basis for the light d.o.f.\ at any given $\mathbf{r}$, but configuration mixing can be permitted at distinct values of $\mathbf{r}$:  $\langle \xi_j (\mathbf{r}') | \xi_i (\mathbf{r^{\vphantom\prime}}) \rangle \neq 0$ in general, even for $j \neq i$.  The full BO approximation then consists of two assumptions: $i$) the light d.o.f.'s in a given ($i^{\rm th}$) eigenstate instantaneously (\textit{adiabatically}) adapt to small changes $\mathbf{r}' \!\! \neq \! \mathbf{r}$ in the heavy-source separation: $\langle \xi_i (\mathbf{r}') | \xi_i (\mathbf{r^{\vphantom\prime}} ) \rangle \approx 1$ (the \textit{adiabatic approximation}); and $ii$) at comparable $\mathbf{r}, \mathbf{r}'$ values, distinct light-field ei\-genstates do not appreciably mix: $\langle \xi_j (\mathbf{r}') | \xi_i (\mathbf{r}) \rangle \approx 0$ for $j \neq i$, the \textit{single-channel approximation}.

The rigorous generalization of the BO approximation to allow for the lifting of these assumptions is called the \textit{diabatic formalism}~\cite{Baer:2006}.  One introduces a free parameter $\mathbf{r}_0$ and rewrites the expansion of the solution Eq.~(\ref{eq:AdExp}) as:
\begin{equation} 
\label{eq:DiaExp}
|\psi \rangle = \sum_{i} \int d\mathbf{r}' \tilde \psi_i
(\mathbf{r}' \! , \mathbf{r}_0) \: |\mathbf{r}' \rangle \:
|\xi_i(\mathbf{r}_0) \rangle .
\end{equation}
Exploiting the completeness of the basis $\{ |\xi_i(\mathbf{r}_0) \rangle \}$ for any specific value $\mathbf{r}_0$, inserting the expansion Eq.~(\ref{eq:DiaExp}) into the \schro equation for the Hamiltonian Eq.~(\ref{eq:SepHam}), and projecting onto $\langle \xi_j(\mathbf{r}_0) |$, one obtains:
\begin{equation}\label{eq:DiaSchro}
\sum_{i} \left[ - \frac{\hbar^2}{2 \mu_{i}} \de_{ji}  \nabla ^2 +
V_{ji}(\mathbf{r},\mathbf{r}_0)- E^{\vphantom \dagger}_i \de_{ji} \right] \! \tilde \psi_i (\mathbf{r},\mathbf{r}_0) = 0.
\end{equation}

The key development is the introduction of the \textit{diabatic potential matrix} $V_{ji}$, defined as:
\begin{equation}
V_{ji}(\mathbf{r},\mathbf{r}_0) \equiv \langle \xi_j (\mathbf{r}_0)|
H_{\rm light} |\xi_i(\mathbf{r}_0) \rangle.
\end{equation}
The parameter $\mathbf{r}_0$ may be chosen as any source separation that {gives a value of energy lying far from a potential-energy level crossing, in which case the states $|\xi_i(\mathbf{r}_0) \rangle$ are unambiguously identified with pure, unmixed configurations identifiable with a single value of $i$.  Meanwhile, $H_{\rm light}$ still references the original source separation $\mathbf{r}$.

Starting with an initial configuration of unmixed $\de \bde$ states and then introducing di-meson states, the diagonal elements of $V$ represent the static light-field energy $V_{\de \bde}$ associated with a pure $\de \bde$ state ($V_{11}$), followed by the potentials $V_{j+1,j+1} = V_{M_1 \bbar M_2}^{(j)}$, $j = 1, 2, \ldots , N$, of the $N$ corresponding di-meson thresholds $(M_1, \bbar M_2)^{(j)}$.  The explicit form of the potential matrix then reads: 
\begin{equation} \label{eq:FullV}
\text V=
\begin{pmatrix}
V_{\de \bde}(\mathbf{r}) & V_{\rm mix}^{(1)}(\mathbf{r})  & \cdots &
V_{\rm mix \vphantom{\bbar M_2}}^{(N)}(\mathbf{r}) \\
V_{\rm mix}^{(1)}(\mathbf{r}) & 
V_{M_1 \bbar M_2}^{(1)}(\mathbf r) &
&
\\
\vdots
& & \ddots \\
V_{\rm mix \vphantom{\bbar M_2}}^{(N)}(\mathbf{r}) & & &
V_{M_1 \bbar M_2}^{(N)}(\mathbf r) \\
\end{pmatrix}.
\end{equation}
Note that we neglect direct mixing terms between any two di-meson configurations by setting the suppressed elements to zero.  Furthermore, for simplicity we set each pure di-meson energy to equal the free energy of the state:
\begin{equation}
V_{M_1 \bbar M_2}^{(j)}(\mathbf r) \to T_{M_1 \bbar M_2}^{(j)} = M_1^{(j)} + M_2^{(j)} \, ,
\end{equation}
although one could of course introduce explicit direct interactions between different di-meson configurations $(j)$, or between the two mesons $(M_1^{(j)} , \bbar M_2)^{(j)}$ within any such configuration.

\section{Distinct Diabatic Thresholds} \label{sec:MultiLevel}

Until very recently, all applications of the diabatic approach to heavy-quark hadrons have assumed that each di-meson channel coupling to the elementary ($Q\bar Q$ or $\de \bde$) state has the same functional form and the same coupling.  In the $\de \bde$ example, 
\begin{equation} \label{eq:Mixpot}
|V_{\rm mix}^{(i)} (r)| = \frac{\Delta}{2}
\exp \! \left\{ -\frac 1 2 \frac{\left[
V^{\vphantom\dagger}_{\de \bde}(r) -
T_{M_1 \bbar M_2 }^{(i)} \right]^2}{\Lambda^2} \right\} ,
\end{equation}
specifically using the same value of $\Delta$ and $\Lambda$ for all channels.  Of course, this {\it Ansatz\/} falls short of elementary expectations even in the heavy-quark limit, where channels composed of various hadron pairs ({\it e.g.}, $D\bar D$ {\it vs.}\ $D \bar D^*$) must differ not only in the mass of the channel (incorporated through $T_{M_1 \bbar M_2 }^{(i)}$), but also in the spin states of the component hadrons.  The formalism for implementing this improvement was developed in Ref.~\cite{Bruschini:2023zkb}; and while its original form refers to $Q \bar Q$ states for which heavy-quark $CP$ is a good quantum number, it can also be applied to $QQ$ states such as $T_{cc}^+$ for which only the $P$ eigenvalue of the heavy-quark pair is a good quantum number.

One of the essential ingredients for incorporating spin dependence into the diabatic formalism is the specification of the overlap of the elementary state with di-meson interpolating operators carrying the same quantum numbers.  In practice, these overlaps are obtained via a Fierz reordering of $(\bar Q \Gamma_1 q)(\bar q \, \Gamma_2 Q)$ operators, where $\Gamma_i$ are Dirac structures, into the form $(\bar Q \Gamma^\prime_1 Q)(\bar q \, \Gamma^\prime_2 q)$.  In the original analysis of Ref.~\cite{Bruschini:2023zkb}, the most interesting $(\bar Q \Gamma_1 q)(Q \Gamma_2 \bar q)$ operators are those that have the same quantum numbers as pure $(\bar Q \Gamma^\prime_1 Q)$ configurations, and hence mix with heavy quarkonium.  On the other hand, $\de \bde$ configurations, already possessing $(Q \Gamma_1^{\prime\prime} q)(\bar Q \Gamma_2^{\prime\prime} \bar q)$ structure, present additional opportunities for nonzero overlaps with $(\bar Q \Gamma_1 q)(\bar q \, \Gamma_2 Q)$ operators~\cite{Lebed:2024}.

In the present case of open heavy flavor, and specifically for the $1^+$ state $T_{cc}^+$, we require a Fierz reordering of the simplest $1^+$ current $J^\mu$ one can construct for $(cc)(\bar u \bar d)$.  The explicit form of this operator, as well as its decomposition in terms of $1^+$ $(\bar u c)(\bar d c)$ operators, first appeared in Ref.~\cite{Xin:2021wcr}, and is presented explicitly in Appendix~A.  Significantly, although the current $J^\mu$ is not a scalar, the derivation of its Fierz reordering directly uses the original Fierz reordering theorem in its proof.\footnote{We thank Z.-G.~Wang for this insight.}  The relevant current and the subset of its couplings to interpolating operators with the quantum numbers of a $(0^-)(1^-)$ di-meson pair, as seen from Eq.~(\ref{eq:Fierz}), reads:
\begin{eqnarray}
J^\mu (x) & = & \epsilon^{ijk} \epsilon^{imn} \bar Q^c_j \gamma^\mu Q_k \, \frac{1}{\sqrt{2}} \left( \bar u_m \, i\gamma^5 d^c_n - \bar d_m \, i\gamma^5 u^c_n \right) \nonumber \\
& \supset & -\frac{1}{\sqrt{2}} (\bar u i \gamma^5 Q) (\bar d \gamma^\mu Q) + \frac{1}{\sqrt{2}} (\bar d i \gamma^5 Q) (\bar u \gamma^\mu Q) . \nonumber \\
\end{eqnarray} 
Noting that the bilinears in the last line of this expression for $Q = c$ provide the simplest interpolating operators for meson pairs $D^0 D^{*+}$ and $D^+ D^{*0}$, respectively, one  may immediately construct the relevant interaction matrix $G(r)$ of Ref.~\cite{Bruschini:2023zkb} [from which, as seen below, the diabatic potential matrix $V(r)$ is derived], where the row/column indices follow the ordering $(cc)(\bar u \bar d)$, $D^0 D^{*+}$, $D^+ D^{*0}$:
\begin{equation} \label{eq:GMatrix} \renewcommand*{\arraystretch}{1.5}
G(r) = \left( \begin{array}{ccc}
V_{(cc)(\bar u \bar d)}   & -\frac{1}{\sqrt{2}} g(r) & +\frac{1}{\sqrt{2}} g(r) \\
-\frac{1}{\sqrt{2}} g(r)  & \Delta_u                       & 0 \\
+\frac{1}{\sqrt{2}} g(r) & 0                                  & \Delta_d
\end{array} \right) \, .
\end{equation}
The function $g(r)$ is defined as the universal string-breaking transition amplitude between the $\de \bde$ and di-meson states.  The latter states are assumed not to have direct transition couplings amongst themselves, and indeed each di-meson state is assumed to behave as a free-particle pair, apart from its $\de$-$\bde$ coupling $V_{\rm mix}$.  Specifically, the constants $\Delta_{u,d}$ are defined as the mass thresholds $\left( m_{D^0} + m_{D^{*+}} \right)$ and $\left( m_{D^+} + m_{D^{*0}} \right)$, respectively, measured with respect to their $I = 0$ combination,
\begin{equation}\label{eq:Tavg}
\frac 1 2 \left( m_{D^0} + m_{D^+} \right) + \frac 1 2 \left( m_{D^{*0}} + m_{D^{*+}} \right) \, .
\end{equation}
which equals $3875.70 \pm 0.05$~MeV\@. Therefore,
\begin{equation}\label{eq:Delta_u}
\Delta_u \equiv \frac 1 2 \left( m_{D^0} - m_{D^+} \right) - \frac 1 2 \left( m_{D^{*0}} - m_{D^{*+}} \right) \equiv -\Delta_d \, ,
\end{equation}
which, using the experimental values given in Footnote~1, is
\begin{equation}
\Delta_u = -0.705 \pm 0.017 \ {\rm MeV} \, ,
\end{equation}
precisely $-\frac 1 2$ times the 1$^{\rm st}$ combination in Eq.~(\ref{eq:TccIsospin}).  The combination $\Delta_u$ breaks both isospin symmetry and heavy-quark spin symmetry.  Thus, parametrically:
\begin{equation}
\Delta_ u \propto \Lambda_{\rm QCD} \, \frac{ m_d - m_u }{ m_c } \, .
\end{equation}

However, the interaction matrix $G(r)$ does not in general equal the full diabatic potential matrix $V(r)$ appearing in the diabatic \schro equation~Eq.~(\ref{eq:DiaSchro}).  The motion of the heavy sources introduced by departures from the static BO limit leads to the mixing of configurations with different heavy-source quantum numbers, as we now discuss.

In the cases investigated in Ref.~\cite{Bruschini:2023zkb}, the light degrees of freedom by themselves are assumed to carry the trivial BO quantum numbers $\Sigma^+_g$, which means a zero-spin projection along the axis of the heavy sources ($\Sigma$), positive parity for reflections in a plane containing this axis ($\epsilon = +$), and light-source $CP$ eigenvalue $\eta = +$, which is denoted by ``$g$''.  In the present case of $QQ\bar u \bar d$, its open-flavor quantum numbers mean that the state is not a $C$ eigenstate, neither in total nor just in its heavy or just in its light components.  Nevertheless, the analysis of Ref.~\cite{Bruschini:2023zkb} remains valid upon identifying $\eta$ eigenvalues for the heavy sources as those of $P$ alone, rather than $CP$.  In particular, the configuration for the heavy degrees of freedom (here, the $QQ$ pair) is specified as $\lambda_\eta$, where $\lambda$ is the projection of heavy-source spin (including sign) along a chosen quantization axis $\hat{\bf z}$, which is not in general the same as the axis $\hat{\bf r}$ defined by the heavy sources in the BO limit.

This mismatch between two natural axis choices---which one may denote as ``space-centered'' versus ``body-centered'' axes, is the origin of the distinction between the interaction matrix $G(r)$ and the diabatic potential matrix $V(r)$.  The relation between them is derived in Ref.~\cite{Bruschini:2023zkb} to be:
\begin{eqnarray}
\lefteqn{V^{\eta, J}_{i, i^\prime, \ell, \ell^\prime} (r) = \sqrt{ (2\ell + 1) (2\ell^\prime + 1) }} & &  \nonumber \\ & \times & \sum_\lambda \left( \begin{array}{ccc} s_i & \ell & J \\ \lambda & 0 & -\lambda \end{array} \right) \left( \begin{array}{ccc} s_{i^\prime} & \ell^\prime & J \\ \lambda & 0 & -\lambda \end{array} \right) G^{\eta, \lambda}_{i, i^\prime} (r) .
\label{eq:VfromG}
\end{eqnarray}
Here, unprimed and primed variables refer, respectively, to initial (row) and final (column) states labeled by $i^{(\prime)}$, and carry spin $s_{i^{(\prime)}}$; the partial wave is labeled by orbital angular momentum $\ell^{(\prime)}$, which combines with $s_{i^{(\prime)}}$ to give the total angular momentum $J$ for the state; and large parentheses indicate Wigner $3j$ symbols.

Under the assumption that the heavy sources interact via a central potential, the usual separation of variables in spherical coordinates also introduces the centrifugal term in the effective potential for specific total $J^P$ quantum numbers:
\begin{equation}
V^{J^P}_{i, i^\prime, \ell, \ell^\prime} (r) = V^{\eta, J}_{i, i^\prime, \ell, \ell^\prime} (r) + \delta_{i, i^\prime} \delta_{\ell, \ell^\prime} \frac{ \ell ( \ell + 1 ) }{2 \mu r^2} ,
\end{equation}
where $\mu$ is the reduced mass of the $(cc)(\bar u \bar d)$ pair.

The specific case of the $J^P = 1^+$ $T_{cc}^+$ is quite trivial in this formalism.  $J=1$, of course; and since we only consider $(cc) (\bar u \bar d)$ and $D D^*$ components, then $i^{(\prime)}$ merely labels the $(cc) (\bar u \bar d)$ state and the two isospin states $D^0 D^{*+}$, $D^+ D^{*0}$ in the order indicated by Eq.~(\ref{eq:GMatrix}), so that $s_1 \! = \! s_2 \! = \! s_3 \! = \!1$; and the $(cc)$ diquark is pure spin-1 with no internal orbital excitation, so that $\lambda = +1, 0, -1$ and $\eta = +$.  The $0^- 0^-$ $D^0 D^+$ state ($\simeq 140$~MeV lower than $T_{cc}^+$) is forbidden from coupling to $1^+$ in any partial wave due to parity conservation; and in this work we neglect the coupling to the combination $D^{*0} D^{*+}$ ($\simeq 140$~MeV higher than $T_{cc}^+$), which allows $s_{i^{(\prime)}} = 0, 1, 2$ (as well as $\lambda = \pm 2$ for the di-meson components), and thus would render Eq.~(\ref{eq:VfromG}) rather more complicated than for the current case.  In fact, while it is straightforward to evaluate the sum in Eq.~(\ref{eq:VfromG}) term by term, in this case the calculation is especially simple because all components have the same value of $s_i$, which implies that the same interaction matrix $G(r) = G^{+, \lambda} (r)$ in Eq.~(\ref{eq:GMatrix}) appears for each of $\lambda = +1, 0, -1$; thus, $G^{+, \lambda}_{i, i^\prime} (r)$ may be taken outside the sum in Eq.~(\ref{eq:VfromG}).  The remaining sum may be rewritten using a completeness relation for $3j$ symbols~\cite{Edmonds:1955fi}:
\begin{eqnarray}
\lefteqn{\sum_{m_1 , m_2} \left( \begin{array}{ccc} j_1 & j_2 & j_3 \\ m_1 & m_2 & m_3 \end{array} \right) \left( \begin{array}{ccc} j_1 & j_2 & j_3^\prime \\ m_1 & m_2 & m_3^\prime \end{array} \right)} & & \nonumber \\
& = & \frac{1}{2j_3 + 1} \, \delta_{j^{\vphantom\prime}_3, j_3^\prime} \, \delta_{m^{\vphantom\prime}_3, m_3^\prime} \, \delta ( j_1, j_2, j_3 ) ,
\end{eqnarray}
where the third $\delta$ factor is the triangle-rule constraint.  Using the invariance of $3j$ symbols under cyclic permutations of their columns~\cite{Edmonds:1955fi} and imposing the constraint $s_{i^\prime} = s_{i^{\vphantom\prime}}$ (unique to this case), the sum in Eq.~(\ref{eq:VfromG}) becomes:
\begin{eqnarray}
\sum_\lambda \left( \begin{array}{ccc} J & s_{i} & \ell \\ -\lambda & \lambda & 0 \end{array} \right) \left( \begin{array}{ccc} J & s_i & \ell^\prime \\ -\lambda & \lambda & 0 \end{array} \right)
& = & \frac{1}{2\ell + 1} \, \delta_{\ell^{\vphantom\prime} \ell^\prime}  \, \delta ( J , s_i , \ell ) ,
\nonumber \\
\end{eqnarray}
and, suppressing $i^{(\prime)}$ indices, Eq.~(\ref{eq:VfromG}) dramatically simplifies to:
\begin{equation}
V^{+, 1}_{\ell, \ell^\prime} (r) = G^{+, \lambda} (r) \, \delta_{\ell^{\vphantom\prime}, \ell^\prime} \, \delta ( 1, 1, \ell) = G(r) \, \delta_{\ell^{\vphantom\prime}, \ell^\prime} \, \delta ( 1, 1, \ell) ,
\end{equation}
so that the interaction matrix $G(r)$ in Eq.~(\ref{eq:GMatrix}) equals the diabatic matrix $V(r)$ in this particular case.  Even though partial waves $\ell = 0, 1, 2$ are allowed, they combine in a very compact manner.  The key ingredient in this simplification is the absence of distinct couplings for different values of $\lambda$, since only one di-meson spin combination (in two isospin channels) occurs; the case of including $D^{*0} D^{*+}$ would be more involved.
 
\section{Analysis} \label{sec:Analysis}

As discussed in Sec.~\ref{sec:Diquark}, the new parameters introduced by the novel nature of a diabatic dynamical-diquark $T^{+}_{cc}$ state are the $V_{\de \bde}$ parameter $R$  [Eq.~(\ref{eq:VforTcc})] indicating the spatial extent of the ``good'' diquark $\bde = (\bar u \bar d)$ as well as its mass $m_\bde = m_{ud}$, and the doubly charmed diquark mass $m_\delta = m_{cc}$.  Additionally, one requires the mixing-potential [Eq.~(\ref{eq:Mixpot})] parameters $\Lambda, \Delta$ introduced by the diabatic formalism~\cite{Lebed:2022vks,Bruschini:2020voj}.  Following the practice of previous works, we set
\begin{equation} \label{eq:rhosigmadef}
    \Lambda=\rho \, \sigma,
\end{equation}
where the string-tension parameter $\sigma$ takes the same value as in Eq.~(\ref{eq:sgmapotparams}), and $\rho$ is a free parameter indicating the width of the mixing potentials.  We first assume that the previous $(\rho,\Delta)$ pair retrieved in Ref.~\cite{Lebed:2023kbm} from fitting to the PDG-averaged $\chi_{c1} (3872)$ mass [Eq.~(\ref{eq:X3872data})],
\begin{eqnarray}\label{eq:rhoDelta}
    \rho & = & 0.165~\rm{fm}, \nonumber \\
    \Delta & = & 0.295~\rm{GeV},
\end{eqnarray}
is appropriate for use in fitting $R$ and $m_{cc}$ to obtain the experimental $T_{cc}^+$ mass.  We later relax this assumption and fit to $(\rho,\Delta)$ as well. 

Our first finding is that the modified potential Eq.~(\ref{eq:VforTcc}), no longer being singular as $r \to 0$, becomes too shallow to reproduce the experimental $T^{+}_{cc}$ mass using the lattice-determined potential parameters of Eq.~(\ref{eq:sgmapotparams}) and any reasonable value for $m_{cc}$.  However, one may note that the eigenvalue $m_{T_{cc}^+}$ has an almost direct dependence on the combination $m_{cc} + V_0$, since $m_{ud}\ll m_{cc}$.  Thus, we introduce an additional offset for the $R > 0$ case, $\Delta_{V_0}=-0.150$~GeV:
\begin{equation}
    V(r) \to V(r) + \Delta_{V_0} ,
\end{equation}
so that, using $V_0$ from Eq.~(\ref{eq:sgmapotparams}),
\begin{equation} \label{eq:V0shift}
    V_0 \to V_0 + \Delta_{V_0} = -0.530~\rm{GeV}.
\end{equation}
With this modification, we find that the pair
\begin{eqnarray}
    R & = & 0.4~\rm{fm}, \nonumber \\
    m_{cc} & =& 3.0260~\rm{GeV},
    \label{eq:Rmccfit}
\end{eqnarray}
produces the experimental value as given in Eq.~(\ref{eq:Tccdata}),
\begin{equation}
    m_{T_{cc}^+}=3874.83~\rm{MeV},
\end{equation}
and the full content of this eigenstate is:
\begin{eqnarray}
    \de \bde & : & \ \, 9.84\% , \nonumber \\
    D^{*+} \! D^{0} & : & 65.57\% , \nonumber \\
    D^{*0} D^{+} & : & 24.59\% . \label{eq:dimesoncontent}
\end{eqnarray}
Since the entire range encompassing $m_{T_{cc}^+}$ and the two thresholds is very narrow ($< \! 1.7~\rm{MeV}$), we see that even the relatively small mass difference $-2\Delta_u$ between the $D^{*+} \! D^{0}$ and $D^{*0} \! D^{+}$ thresholds [Eq.~(\ref{eq:TccIsospin})] is enough to allow $ D^{*+} \! D^{0}$ to dominate.  This clear demonstration of isospin symmetry breaking can also be directly explored by allowing the value of $\Delta_u$ to vary.  In Fig.~\ref{fig:DeltauThresholds}, we show that adjusting $\Delta_u$ from $-0.705$~MeV to 0~MeV brings the difference between $D^{0} D^{*+}$ and $D^{*0} D^{+}$ content to zero quadratically in $\Delta_u$, the two meeting at around 44\%.  Notably, this modification also relaxes the overall dominance of the di-meson thresholds on the state content, allowing for the $\de \bde$ content to rise to a maximum of about $11 \%$, as seen in Fig.~\ref{fig:DeltauDiquark}.  Of course, any change in $\Delta_u$ while holding other parameters fixed [particularly, the isospin-averaged threshold of Eq.~(\ref{eq:Tavg})] directly changes the eigenvalue $m_{T_{cc}^+}$, as seen in Fig.~\ref{fig:DeltauEigs}.

Figure~\ref{eq:Tccdata} also illustrates an interesting effect expected from the diabatic formalism: The resonance mass eigenvalue (here, $m_{T_{cc}^+}$) prefers to stay closer to the lower threshold, rather than reataining a special affinity to the particular threshold ($D^{*+} \! D^0$) to which it is closest at the physical point.  Moreover, the distance of the eigenvalue below the thresholds is maximal (about 700~keV) when isospin breaking vanishes ($\Delta_u = 0$) and the thresholds coincide.  Lastly, at larger values of isospin breaking ($|\Delta_u| > 1.3$~MeV, where the $DD^*$ thresholds differ by more than 2.6~MeV), $m_{T_{cc}^+}$ rises above the lower threshold, and the $T_{cc}^+$ width would be expected to increase dramatically.

Although Figs.~\ref{fig:DeltauThresholds}--\ref{fig:DeltauEigs} are almost perfectly (anti)sym\-metric about $\Delta_u = 0$, one may note asymmetries about this point, especially in the quadratic fits in Fig.~\ref{fig:DeltauThresholds}, which have explicit discontinuities at $\Delta_u=0$.  These jumps originate from the fact that the physical point $\Delta_u=-0.705$ MeV is special: i.e., the mass eigenvalue $m_{T_{cc}^+}$ (and thus the potential parameters) is initially fitted at this value, and only then is $\Delta_u$ varied.  While the quadratic fits of Fig.~\ref{fig:DeltauThresholds} are discontinuous at $\Delta_u=0$, the content for each di-meson component is seen to vary smoothly through this point.  For a more quantitative description of this asymmetry, we fit to the functional form $a\Delta^{2}_u + b\Delta_u + c$ and for each di-meson component, providing the fit parameters $(a,b,c)$ (which have units of MeV$^{-(2,1,0)}$, respectively):
\begin{eqnarray}
    D^{*0} \! D^{+} \; & (\Delta_u < 0): & (-0.0623,-0.3651,0.4289), \nonumber \\
    D^{*0} \! D^{+} \; & (\Delta_u > 0): & (+0.0744,-0.3397,0.4412), \nonumber \\
    D^{0} D^{*+} \; & (\Delta_u < 0): & (+0.0698,+0.3268,0.4404), \nonumber \\
    D^{0} D^{*+} \; & (\Delta_u > 0): & (-0.0569,+0.3507,0.4282).
\end{eqnarray}
    
As suggested above, one may use $m_{T^{+}_{cc}}$ as a fixed starting point, and allow the diabatic parameters $\rho$ and $\Delta$ to be fit as well.  In doing so, $T^{+}_{cc}$ may be used as a laboratory to explore the relationships of the relevant diabatic dynamical-diquark parameters.  Since fitting to the full parameter space of $(R,m_{cc},\Delta_{V_0},\rho,\Delta)$ using the single state $T^{+}_{cc}$ allows many solutions, we have performed multiple experiments by holding some parameters constant and varying others.  Using the determination of $(R,m_{cc})$ from Eq.~(\ref{eq:Rmccfit}), we vary $(\Delta_{V_0},\rho,\Delta)$.  We find that a value of $\Delta_{V_0}=-0.150~\rm{GeV}$ allows for the largest set of $(\rho,\Delta)$ pairs that successfully fit the $m_{T^{+}_{cc}}$ (hence its usage in the initial fit above), with a steep dropoff in the space of suitable $(\rho,\Delta)$ pairs, whether one increases or decreases $\Delta_{V_0}$.  Within the $(\rho,\Delta)$ subspace for this $\Delta_{V_0}$ value, we observe a quadratic relationship between $(\rho,\Delta)$ fit values, as shown in Fig.~\ref{fig:rhodeltaplot}.  Clearly, the ability to freely vary such phenomenologically critical parameters as $\rho, \Delta$ across such a wide range and still obtain equivalent fits shows that a separate, precision determination of the parameters from lattice QCD would have a significant impact on the understanding of near-threshold states, as discussed in Refs.~\cite{Lebed:2022vks,Bruschini:2020voj}.

\begin{figure*}
    \centering
    \includegraphics{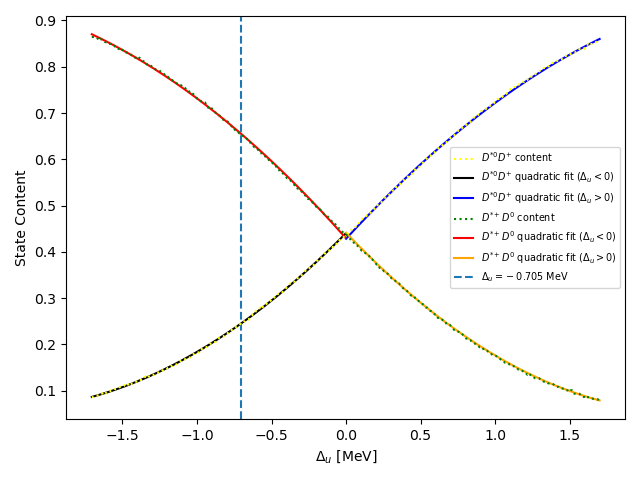}
    \caption{Content of the normalized $T_{cc}^+$ eigenstate associated with free di-meson configurations $D^{*+} \! D^{0}$ and $D^{*0} D^{+}$ as a function of the isospin-breaking parameter $\Delta_u$ [Eq.~(\ref{eq:Delta_u})] for potential parameters given in Eqs.~(\ref{eq:rhoDelta}), (\ref{eq:V0shift}), and (\ref{eq:Rmccfit}).  The isospin-averaged threshold mass [Eq.~(\ref{eq:Tavg})] is held fixed.  Additionally, we fit quadratic curves to each component separately for $\Delta_u < 0$ [which contains the physical point, represented by a dashed vertical line, from which agreement can be seen with the results in Eq.~(\ref{eq:dimesoncontent})] and for $\Delta_u >0$.}
    \label{fig:DeltauThresholds}
\end{figure*}

\begin{figure*}
    \centering
    \includegraphics{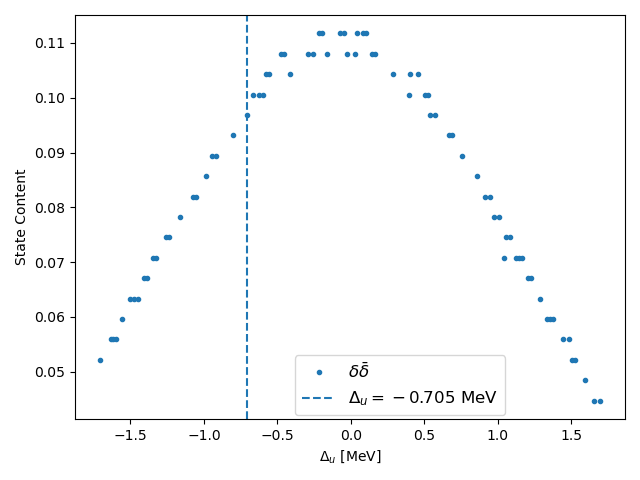}
    \caption{Content of the normalized $T_{cc}^+$ eigenstate associated with the $\de \bde$ configuration as a function of the isospin-breaking parameter $\Delta_u$ [Eq.~(\ref{eq:Delta_u})], using the same potential parameters and isospin-averaged mass as in Fig.~\ref{fig:DeltauThresholds}.  The dashed vertical line again represents the physical point, from which agreement can be seen with the results in Eq.~(\ref{eq:dimesoncontent})].}
    \label{fig:DeltauDiquark}
\end{figure*}

\begin{figure*}
    \centering
    \includegraphics{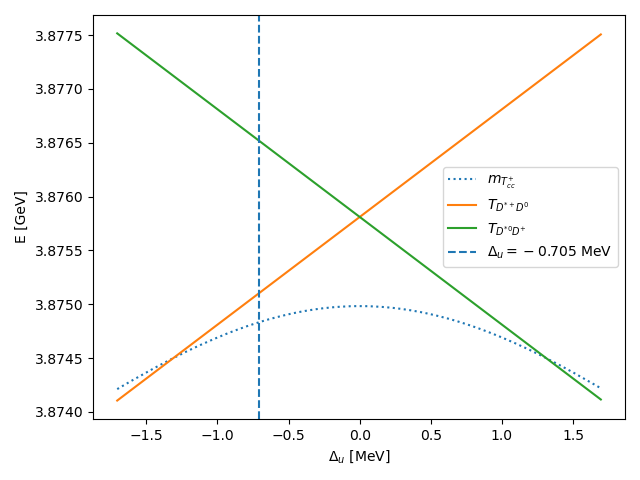}
    \caption{The mass eigenvalue $m_{T^{+}_{cc}}$ as a function of $\Delta_u$, using the same potential parameters and isospin-averaged mass [Eq.~(\ref{eq:Tavg})] as in Figs.~\ref{fig:DeltauThresholds}, \ref{fig:DeltauDiquark}. The di-meson threshold energies are included for reference.  The dashed vertical line again represents the physical point, from which agreement with the measured values in Eqs.~(\ref{eq:Tccdata}) and (\ref{eq:TccIsospin}) can be obtained.}
    \label{fig:DeltauEigs}
\end{figure*}
    
\begin{figure*}
    \centering
    \includegraphics{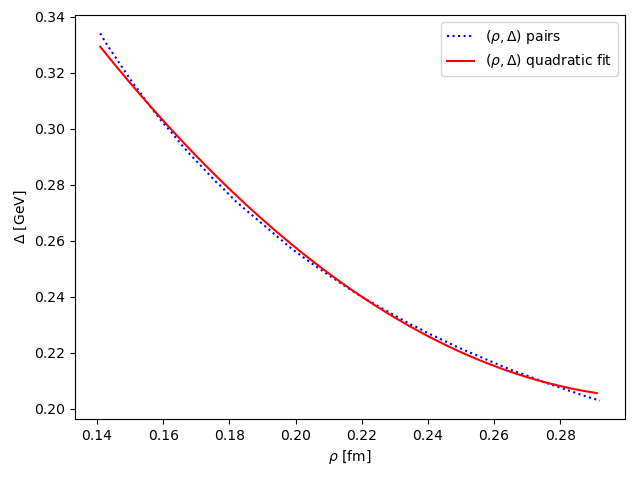}
    \caption{Pairs of diabatic-potential parameters $\rho, \Delta$ [Eqs.~(\ref{eq:Mixpot}), (\ref{eq:rhosigmadef})] that give equivalent fits to the experimental value of $m_{T_{cc}^+}$ [Eq.~(\ref{eq:Tccdata})], fixing $\Delta_{V_0} = -0.150$~MeV [Eq.~(\ref{eq:V0shift})].  Also presented is a quadratic fit to the results.}
    \label{fig:rhodeltaplot}
\end{figure*}

\section{Conclusions} \label{sec:Concl}

We have performed the first study of the open-charm state $T_{cc}^+$ in the diabatic generalization of the Born-Oppenheimer approximation.  In particular, the $T_{cc}^+$ substructure is dominated by its extreme proximity to the $D^{*+} \! D^0$ threshold, but it also lies not much further below the $D^{*0} D^+$ threshold, and in addition permits a $(cc)(\bar u \bar d)$ diquark-antidiquark ($\de \bde$) component.  As expected [and in analogy to a similar situation for $\chi_{c1}(3872)$ with the $D^{*0} \bar D^0 + D^0 \bar D^{*0}$ threshold], we find the mass eigenstate to consist overwhelmingly ($> \! 90\%$) of di-meson components, but the $D^{*0} D^+$ component is nevertheless more than 1/3 as large as the $D^{*+} \! D^0$ component that lies almost atop $m_{T_{cc}^+}$.  Moreover, the elementary $\de \bde$ component persists in both cases at a level of nearly $10\%$, providing a significant short-distance component to the wave function of both $T_{cc}^+$ and $\chi_{c1}(3872)$.

The two states nevertheless differ in key respects.  $\chi_{c1}(3872)$ has several known decay channels to charmonium, which of course cannot occur for the open-charm $T_{cc}^+$.  But the two $D^* \! D$ (or $D^* \! \bar D$) isospin-partner decay modes are much more closely spaced for $T_{cc}^+$ than for $\chi_{c1}(3872)$, making $T_{cc}^+$ a superior laboratory for studying isospin breaking in 4-quark states.  We have used this feature to compute how all 3 $T_{cc}^+$ components: $D^{*+} \! D^0$, $D^{*0} D^+$, and $\de \bde$, change as the isospin-breaking mass difference $m^{\vphantom\dagger}_{D^{*+} \! D^0} - m^{\vphantom\dagger}_{D^{*0} D^+}$ is adjusted.  We find that the $\de \bde$ component is fairly stable (between $5$--$11\%$) over a large range of this difference, with its maximum occurring close to the isospin-symmetric point.  Since we have fixed the measured value of $m_{T_{cc}^+}$ at the specific physical value of isospin breaking, a small$^{\vphantom\dagger}$ asymmetry corresponding to exchanging the $D^{*+} \! D^0$ and $D^{*0} D^+$ components arises.  Nevertheless, we conclude that the dominant parameter determining both the variation of the $\de \bde$ state content and the eigenvalue $m_{T_{cc}^+}$ appears to be the magnitude of the isospin-breaking mass difference.

We have noted above that the particular modeling of the $T_{cc}^+$ system in this paper requires a very simple diabatic potential matrix, because we neglected the more distant $D^{*+} \! D^{*0}$ threshold.  Incorporating this threshold, which requires the use of a more involved spin-dependent formalism, is one key direction of future research.  Indeed, properly including spin dependence in the diabatic potential matrix has not yet been performed in the hidden-charm sector of this model, and constitutes its own set of projects still to be carried out.  We have also noted that, if the ``good'' ($\bar u \bar d$) diquark is in fact an essential component of $T_{cc}^+$, then this state likely has no nearby multiplet partners, unlike the case of the hidden-charm multiplet containing $\chi_{c1}(3872)$, $Z_c(3900)$, $Z_c(4020)$, and presumably several other states.  In several ways, the state $T_{cc}^+$ provides an ideal laboratory in which to test different diabatic hypotheses with a minimum of complications.

\appendix
\section{Fierz Reordering of Isoscalar Diquark-Antidiquark Vector Current}

For an arbitrary fermion field $\psi$, define its charge-conjugate form in the usual manner:
\begin{equation}
\psi^c (x) \equiv C {\bar \psi}^T \, ,
\end{equation}
where $C$ is the charge-conjugation Dirac matrix, which has the properties
\begin{equation}
C \gamma^\mu C^{-1} = -\gamma^{\mu \, T} \, , \ \ C^{-1} = C^T = C^\dagger = -C = -C^* \, . 
\end{equation}
From these relations also follows:
\begin{equation}
{\bar \psi}^c (x) = \psi^T \! (x) \, C \, .
\end{equation}

The Hermitian interpolating operator for the ``good'' (spin-0, isoscalar, color-triplet) light diquark ($\bar u \bar d$) is~\cite{Jaffe:2004ph}:
\begin{equation} \label{eq:good}
\epsilon_{ijk} \frac{1}{\sqrt{2}} \left( \bar u_j \, i\gamma^5 d^c_k - \bar d_j \, i\gamma^5 u^c_k \right) ,
\end{equation}
where $i,j,k$ are color indices.

Let $J^\mu (x)$ be the simplest current for creating a $J^P = 1^+$ state with flavor quantum numbers $QQ \bar u \bar d$, with the light quarks being in the ``good'' diquark configuration of Eq.~(\ref{eq:good}).  Then it is a straightforward (but lengthy) exercise to apply the conventional Fierz reordering to the $Q\bar q$ and $q^c {\bar Q}^c$ field pairs to obtain:

\begin{eqnarray}
\lefteqn{J^\mu (x) = \epsilon^{ijk} \epsilon^{imn} \bar Q^c_j \gamma^\mu Q_k \, \frac{1}{\sqrt{2}} \left( \bar u_m \, i\gamma^5 d^c_n - \bar d_m \, i\gamma^5 u^c_n \right)} & &  \nonumber \\
& = & -\frac{1}{\sqrt{2}} (\bar u i \gamma^5 Q) (\bar d \gamma^\mu Q) + \frac{1}{\sqrt{2}} (\bar d i \gamma^5 Q) (\bar u \gamma^\mu Q) \nonumber \\
& & + \frac{i}{\sqrt{2}} (\bar u Q) (\bar d \gamma^\mu \gamma^5 Q) - \frac{i}{\sqrt{2}} (\bar d Q) (\bar u \gamma^\mu \gamma^5 Q) \nonumber \\
& & -\frac{1}{\sqrt{2}} (\bar u \, \sigma^{\mu \nu} Q) ( \bar d \gamma_\nu \gamma^5 Q) +\frac{1}{\sqrt{2}} (\bar d \, \sigma^{\mu \nu} Q) ( \bar u \gamma_\nu \gamma^5 Q) \nonumber \\
& & +\frac{1}{\sqrt{2}} (\bar u \, \sigma^{\mu \nu} \gamma^5 Q) ( \bar d \gamma_\nu Q) -\frac{1}{\sqrt{2}} (\bar d \, \sigma^{\mu \nu} \gamma^5 Q) ( \bar u \gamma_\nu Q) \, .
\nonumber \\ \label{eq:Fierz} 
\end{eqnarray}

The tetraquark state thus couples naturally not only to di-meson pairs with quantum numbers $(0^-)(1^-)$, but also to $(0^+)(1^+)$, $(2^+)(1^+)$, and $(2^-)(1^-)$.  Moreover, the (pseudo-)tensor bilinear operators in the final line of Eq.~(\ref{eq:Fierz}) can also serve as interpolating operators for vector particles, which is how the overlap with $(1^-)(1^-)$ pairs (such as $D^* D^*$ for $T_{cc}^+$) occurs in this formalism.

Analogous Fierz reorderings between diquark-anti\-di\-quark and di-meson scalar operators, as would appear in Lagrangians, are presented in Appendix~A of Ref.~\cite{Buballa:2003qv}.

\begin{acknowledgments}
We thank R.~Bruschini, E.~Braaten, and Z.-G.~Wang for insightful comments.  This work was supported by the National Science Foundation (NSF) under Grants No.\ PHY-1803912 and PHY-2110278.
\end{acknowledgments}

\bibliographystyle{apsrev4-2}
\bibliography{diquark}
\end{document}